# Evaluating data-driven background ensembles covariances from Graphcast: a case study for Hurricane Lee (2023)

Zhihong Chen,[a] Xuguang Wang,[a,b]

[a] *University of Oklahoma, Norman, OK*

[b] *Consortium for Advanced Data Assimilation Research and Education*

*Corresponding author*: Zhihong Chen, zchen@ou.edu

ABSTRACT

Short-term background ensemble covariances (BEC) are crucial for ensemble-based data assimilation (DA). However, limited studies so far have examined the fidelity of the cost-effective data-driven model in producing the short-term BEC for hurricane data assimilation.

In this study, we evaluate the background ensemble spread and correlations from GraphCast against those of GEFS for Hurricane Lee (2023) during both its intensification and non-intensification phases. Specifically, the BEC in the hurricane vortex, the hurricane environment, and the vortex–environment interactions are examined.

Within the hurricane vortex, the background ensemble of GraphCast is less dispersive than GEFS. The two models agree well on the background ensemble correlations that are tied to the primary circulation but show a larger correlation difference associated with the secondary circulation, indicating the two models represent unbalanced and diabatic processes differently. In the hurricane environment, binned univariable correlations show linear relationships between the two models, with a weaker horizontal geopotential height correlation in GraphCast. GraphCast also shows a reduced spread and a flatter empirical orthogonal function spectrum of the 500 hPa geopotential height background ensemble, with more perturbation growth distributed to smaller-scale features such as shortwaves. For the vortex–environment interaction, the two models produce close background ensemble correlation patterns for Lee's track but differ more for intensity.

Overall, GraphCast can produce broadly consistent short-term BEC for Hurricane Lee compared to those of GEFS. However, systematic difference exists in certain variables, scales, and processes, suggesting the need to further investigate its fidelity in a cycled hurricane DA context.

## 1. Introduction

Recent studies have shown that data-driven weather models can obtain forecast skills comparable to the leading physics-based numerical weather prediction (NWP) systems in various metrics while running orders of magnitude faster. For example, deterministic data-driven models such as FourCastNet, Pangu-Weather, GraphCast, and FuXi have shown promising skills for medium-range forecasts (Pathak et al., 2023; Bi et al., 2023; Lam et al., 2023; Chen et al., 2023). Diffusion approaches, such as SEEDS and GenCast, which sample from learned probabilistic distributions, demonstrate the ability to produce hundreds to thousands of ensemble members with calibrated ensemble spread at low computational cost (Li et al., 2024; Price et al., 2025). In the field of ensemble-based data assimilation (DA), various efforts have been made to mitigate the sampling error due to the use of a limited number of short-term background ensemble members because of the computational cost constraints (e.g. Hamill et al., 2001; Hunt et al., 2007; Wang et al., 2021). Data-driven models could change the cost curve for ensemble-based data assimilation. If the short-term forecasts and their derived background ensemble covariances (BEC) from such data-driven models are reliable, very large ensembles would become practical at modest cost, reducing the sampling error while preserving the flow-dependent error covariance structure. But the question is–are these data-driven models ready to surrogate the short-term background forecasts?

Previous studies revealed concerns in a few aspects. Most deterministic data-driven models optimize a squared-error loss, which targets conditional means and suppresses variance, making the model under-dispersive (Murphy, 2022, Lam et al., 2023). Curriculum learning and fine-tuning on the next several forecast steps reduce the mean squared error (MSE) in longer lead time, but sacrifice the model's optimality on the short-term forecast (Pathak et al., 2023, Lam et al., 2023, Chen et al., 2023). Examinations of the tangent–linear and adjoint models of data-driven weather forecast models show non-physical sensitivities and model imbalances in jet dynamics and wave propagation (Baño-Medina et al., 2025, Tian et al., 2026). Additionally, previous work has shown that data-driven models may fail to capture the growth of perturbation if the magnitude of the initial perturbation is too small, whereas physics-informed perturbation schemes help produce realistic error-growth rates and directions (Selz and Craig, 2023; Pu et al., 2025).

These aforementioned issues may cause challenges in using data-driven weather models to produce the surrogate background ensemble for ensemble-based data assimilation. Kotsuki et al. (2025) demonstrated that careful tuning of localization and inflation parameters is needed for a low-resolution data-driven global model ClimaX when it is used to cycle with the local ensemble transform Kalman filter (LETKF). Another study assimilating observed surface pressure into several data-driven models revealed that accumulation of small-scale noise can lead to filter divergence, unless the spectral filtering was applied (Slivinski et al., 2025). These results point to the need for a careful evaluation of the BEC produced by the data-driven global weather models toward their ensemble data assimilation application.

Hurricanes are high-impact weather systems that involve complex multiscale dynamic and thermodynamic interactions. Within the hurricane vortex, deep convection and latent heat release drive the primary and secondary circulations and are tightly coupled with eyewall and rainband dynamics (e.g. Shapiro and Willoughby 1982, Emanuel 1986, Stern and Nolan 2009, Zhang et al., 2011). At the synoptic scale, subtropical ridges, midlatitude troughs, and the ITCZ modulate the tracks and recurvature of hurricanes (Chan and Garry, 1982; Frank and Ritchie, 2001, Hsiao et al., 2015). Recent studies demonstrate the need to use multiscale methods to treat the sampling errors of the BEC to improve the hurricane analysis and subsequent forecast (Lu and Wang 2024). To the authors' best knowledge, there has been no study that evaluates the BEC estimated with data-driven ensembles on hurricanes.

This study aims to fill the gap by answering the following research questions using a case study approach: Can a data-driven model produce short-term BEC related to hurricanes with realistic amplitude and structure across variables and scales? As the first step of the investigation, we compare the 6-hour BEC of a data-driven model, GraphCast (Lam et al., 2023), to the 6-hour BEC of a physics-based ensemble prediction system, the Global Ensemble Forecast System (GEFS[WX2.1]) (Zhou et al., 2022), for Hurricane Lee (2023). Given that the 6-hour GEFS ensembles have been extensively used in cycled operational Global Data Assimilation System (GDAS) (e.g. Whitaker et al., 2008, Wang et al. 2013, Hamill et al., 2022, Ye et al., 2023), the similarities and differences between the BECs of GraphCast and GEFS provide a reference to assess the potential of data-driven models in providing surrogate background ensembles for hurricane data assimilation. Section 2 describes the experiment designs. Section 3 examines BEC within the hurricane vortex. Section 4 examines BEC in large scale hurricane environment using principal component analysis. Section 5 diagnoses BEC in simulating the interaction between the hurricane vortex and hurricane environment. And Section 6 summarizes the results.

## 2. Experiment Design

*a. Case Selection*

Hurricane Lee (2023) is chosen as the test case for this study due to its high impacts and complex intensification processes (Blake and Nepaul, 2024). Lee formed from an African easterly wave on 5 September 2023, rapidly intensified for 70 kt in 24 hours from 7 September 2023, and peaked as a Category 5 hurricane at 0600 UTC 8 September 2023. It weakened due to the increasing southwest shear and fell below major hurricane strength on 9 September 2023. On 10-11 September 2023, the eyewall reorganized, and Lee re-intensified to 105 kt, while undergoing eyewall-replacement cycles that expanded the wind field. Lee then gradually weakened, moved northward, made landfall near Long Island, and produced destructive storm surge and widespread power outages in New Brunswick and Prince Edward Island. In this study, two phases are examined: a non-intensification (NON-IN) stage at 0600 UTC 9 September 2023 and an intensification (IN) stage with eyewall-replacement at 1800 UTC 10 September 2023 (Fig. 1). Comparison between the two phases provides a more complete view of how GraphCast estimates BECs associated with different hurricane dynamic regimes.

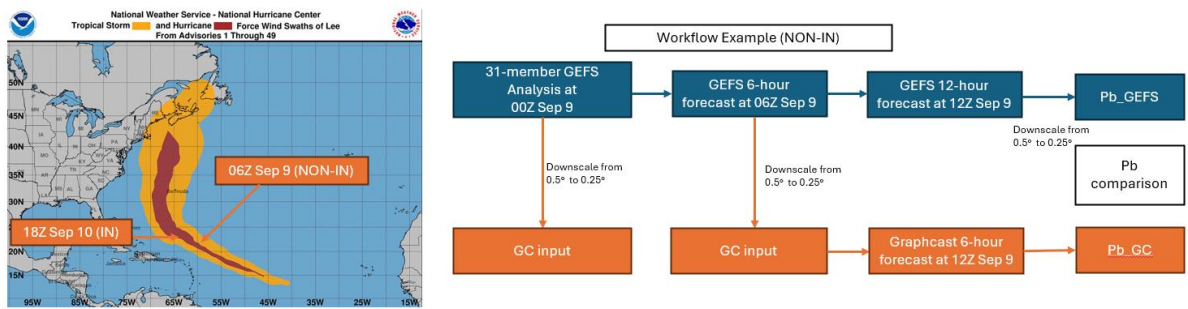

**Figure 1.** Left: the track of Hurricane Lee (National Hurricane Center, 2023), orange arrows point to the two analysis times: 0600 UTC 9 September 2023 9 (NON-IN) and 1800 UTC 9 September 2023 (IN). Right: an illustration of the experiment design.

*b. Experiment Design*

GraphCast is selected for this study. GraphCast features a graph neural network on a spherical mesh trained with a deterministic loss function and has demonstrated promising medium-range forecast skill at 0.25° with low inference cost (Lam et al., 2023). GraphCast has been evaluated by NOAA within the Experimental AI Global and Limited-area Ensemble forecast system (EAGLE) project, and community tools exist to initialize GraphCast directly from the GFS or GEFS analyses, enabling member-matched experiments with GEFS initial conditions (NOAA EPIC, 2025; Radford, 2024). GraphCast needs two consecutive steps of initial conditions that are 6 hours apart to initialize the forecasts. The GEFS analysis and its 6-hour forecasts of 0.5° resolution are downscaled to 0.25° before input into GraphCast to initialize the 6-hour forecasts. To ensure comparison at the same forecast valid time, these GraphCast forecasts are compared with the GEFS with a 12-hour lead time. The BEC of the hurricane vortex, the hurricane environments, and the vortex–environment interactions are then calculated and compared between GEFS and GraphCast. The workflow of forecast initialization and BEC comparison is shown in Fig. 1. Note that this study does not perform a cycling DA experiment. Instead, this experiment design enforces that both systems share identical initial perturbations from the GEFS analysis. This design facilitates a controlled comparison of the BECs, isolating the choice of model, emulator versus physics-based, as the primary factor.

*c. Definition of Hurricane Vortex and Hurricane Environment*

Following the definition of scales in Lu and Wang (2024), the hurricane vortex is defined as all grid points within 500km of the hurricane center. Grids are binned into 10 radial bins with a 50 km increment. Each variable among geopotential height (GH), temperature (T), specific humidity (Q), azimuthal wind (AW), and radial wind (RW) at each of the 13 vertical levels from 1000 hPa to 100 hPa) is averaged within the bin. This yields 650 (10 radial bins × 13 levels × 5 variables) elements, which is the length of the state vector.

The hurricane environment is a 6000 × 6000 km square centered on the mean hurricane center location in the GEFS 6-hour forecasts, but excluding the area of the hurricane vortex. Effectively, a 3000 km radial cutoff (Lu and Wang, 2024) was applied to the BEC of the hurricane environment to filter out the distant spurious correlations. Given the dimension of the BEC in this domain for EOF calculation, 500 hPa GH is selected as a representative field for some the diagnosis in the hurricane environment section.

*d. Background Ensemble Variance, Correlation and Correlation Tendencies*

To examine BEC, its variance and correlation are calculated. Additionally, the 6-hour background ensemble correlation tendency is also examined and compared between GraphCast and GEFS. The 6-hour correlation tendency is defined as the correlation in the 6-hour forecast ($Corr\_t6$) minus the correlation in the initial conditions ($Corr\_t0$):

$$dCorr = Corr_{t6} - Corr_{t0}$$

When the magnitude of the correlation in initial condition | $Corr\_t0$ | is large and the 6-h changes are small, the two models may appear to have similar $Corr\_t6$ simply because both retain the strong correlation structure already present at initialization, even if they produce different incremental changes over the 6-h forecast. We therefore examine the correlation tendency $dCorr$ to explicitly examine the correlation evolution induced by the two models.

## 3. Results

*a. Hurricane Vortex*

The primary and secondary circulations within the hurricane vortex are tightly coupled with the thermodynamic structure and development of the hurricanes. Accurate representation of the BEC of wind, temperature and moisture within the vortex is essential for hurricane DA and forecast. In this section, BECs from the GraphCast and the physics-based model are compared.

Background ensemble spread inside the hurricane vortex is examined first. Figure 2 shows the ratio of GraphCast ensemble standard deviation (STD) to GEFS ensemble STD for NON-IN and IN. In both phases, the ratio is less than one for all variables, suggesting GraphCast is systematically less dispersive compared to GEFS within the hurricane vortex. This result is consistent with previous studies showing that data-driven models trained with a deterministic loss function tend to be under-dispersive (Lam et al., 2023; Subich et al., 2025). From NON-IN to IN, the ratios further decrease, with GH ratio dropping most from about 0.80 to 0.62. It is hypothesized that the perturbation growth associated with the stronger influence of deep convection and more complex subgrid processes during intensification may be less represented in the GraphCast trained with a deterministic loss than in the GEFS system. Among the five variables, RW has the smallest spread ratio for both phases (0.59 and 0.52), suggesting that the growth of perturbation associated with the secondary circulation is significantly smaller in GraphCast compared to GEFS.

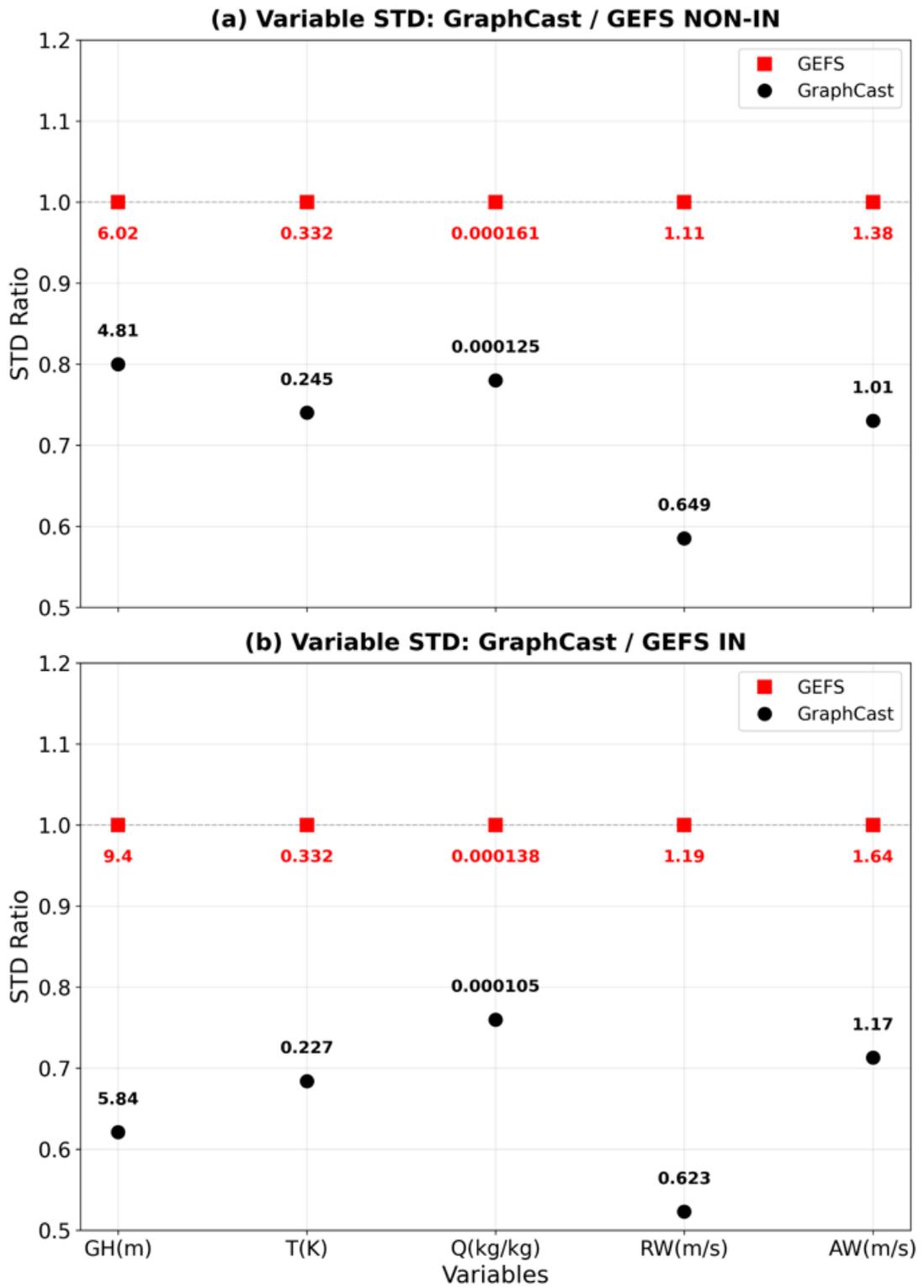

**Figure 2.** Ratio of GC to GEFS ensemble standard deviation within the hurricane vortex for the NON-IN (a) and IN (b) phases for geopotential height (GH), temperature (T), specific humidity (Q), radial wind (RW), and azimuthal wind (AW). Red squares denote GEFS reference, black circles denote the GraphCast ratio, values of the spread are shown above the squares and circles. The dashed line (ratio = 1) denotes equal spread between GraphCast and GEFS.

The magnitude of background ensemble correlation and correlation tendency of the hurricane vortex are then examined for 15 variable combinations in Fig. 3. Figure 3a and 3c compare the 6-h mean absolute correlations of each variable pair between GraphCast and GEFS for NON-IN and IN. Most points lie close to the $y = x$ line, indicating the magnitude of the hurricane inner-core background ensemble correlations of GraphCast is in general in agreement with that from GEFS. Fig. 3b and 3d relate the mean absolute correlation tendency over 6 hours $|dCorr|$ to the mean absolute correlation in the initial condition. For both the GraphCast and GEFS ensembles, pairs with larger absolute initial correlation values change less over the 6-hour period. GH-involved pairs (e.g., GH–GH, GH–AW) with stronger correlation in the initial condition display smaller $|dCorr|$, while pairs involving RW, T, and Q with weaker correlations in initial conditions show larger $|dCorr|$. This result agrees with previous studies that balanced relationships are maintained by the model over short lead times, while unbalanced and diabatic processes generate faster, less predictable error growth and thus larger correlation adjustments (Fisher, 2003, Zhang et al., 2007). For almost every variable pair, GraphCast consistently shows larger $|dCorr|$ than GEFS. This difference of $|dCorr|$ is larger in the IN than in NON-IN, suggesting Grapchast forecasts more rapid spatial correlation evolution compared to GEFS, especially during the intensification.

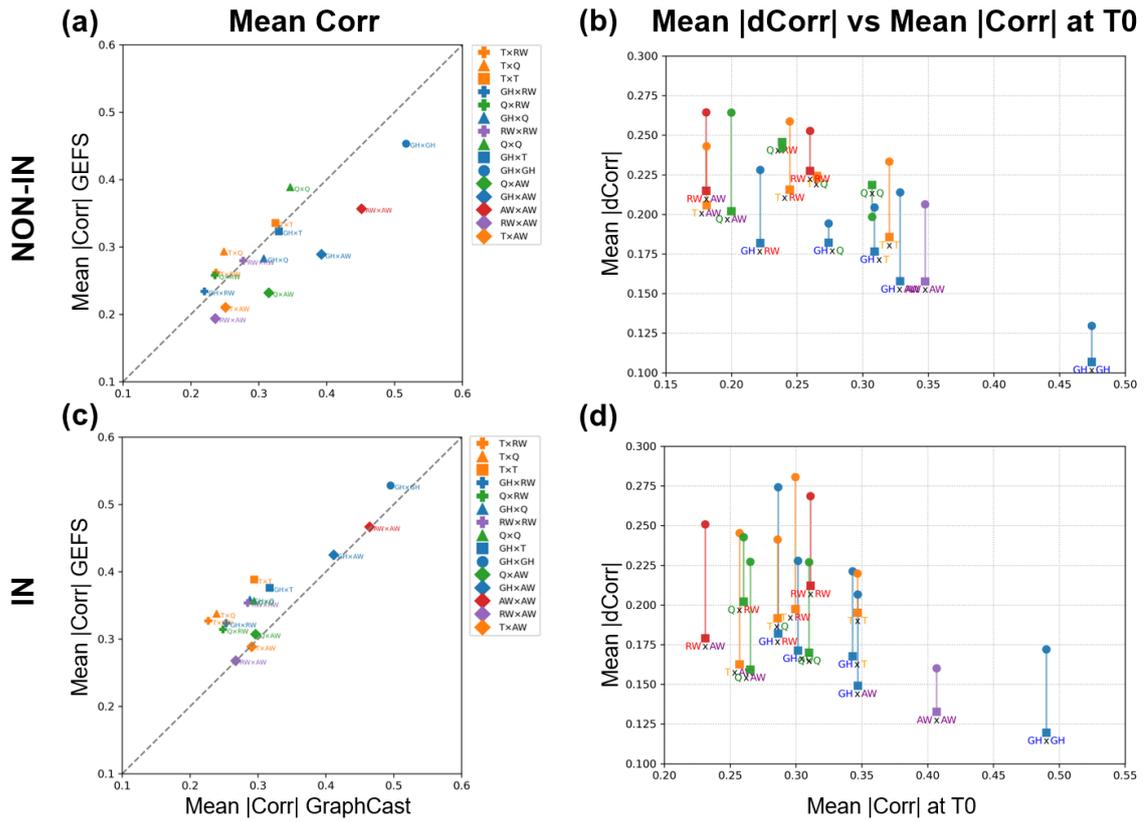

**Figure 3.** (a,c) Mean absolute 6-h correlations $|Corr|$ for 15 variable pair from GC (x-axis) and GEFS (y-axis) for NON-IN and IN; (b,d) Mean absolute correlation tendency $|dCorr|$ (y-axis) versus mean absolute correlation in the initial condition $|Corr\_t0|$ (x-axis). For each variable pair, round dots denote GraphCast and square dots denote GEFS, connected by a vertical line. Variable pair labels are annotated next to each marker.

Larger $|dCorr|$ in GraphCast does not necessarily imply an advantage. Whether the background correlation and correlation tendency of GraphCast are physically realistic needs further diagnosis. To filter out small-scale noise and average out individual grid-point mismahurricanehes, a binned regression approach (Wang and Bishop 2003) is applied to compare the correlations of GraphCast and GEFS. Univariate correlations of variables are examined and three representative variables (AW–AW, GH–GH, Q–Q) are shown in Fig. 4. For AW (Fig. 4 a, d, g, j), the correlation and correlation tendency of GraphCast and GEFS are strongly linearly related ($R^2 > 0.95$), indicating that the correlation structures for the primary circulation of the two models are broadly consistent. For GH, the correlations (Fig. 4 b, h) match well (binned scatters are close to $y = x$) because GH–GH has the strongest correlations in the initial conditions (Fig. 3b and 3d). However, the correlation tendencies (Fig. 4 e, k) show a weak linear relationship ($R^2 < 0.5$), indicating that the two ensembles differ substantially in how GH correlations evolve over 6 hours. This is consistent with Fig. 3b and 3d, where GH–GH is among the smallest correlation tendencies in both systems. In other words, small signals are harder to capture robustly. This mismatched correlation tendency could be a problem for DA. Through continuous DA cycling, non-physical correlation tendency may accumulate and lead to filter divergence, which is observed in Slivinski et al. (2025). For Q, the correlations and tendencies are strongly linearly correlated between GraphCast and GEFS in NON-IN (Fig. 4 c, f, $R^2 > 0.95$), but the linear relationship degrades during IN (Fig. 4i, j), suggesting GraphCast may handle deep convection occurring in the hurricane intensification more differently compared to GEFS.

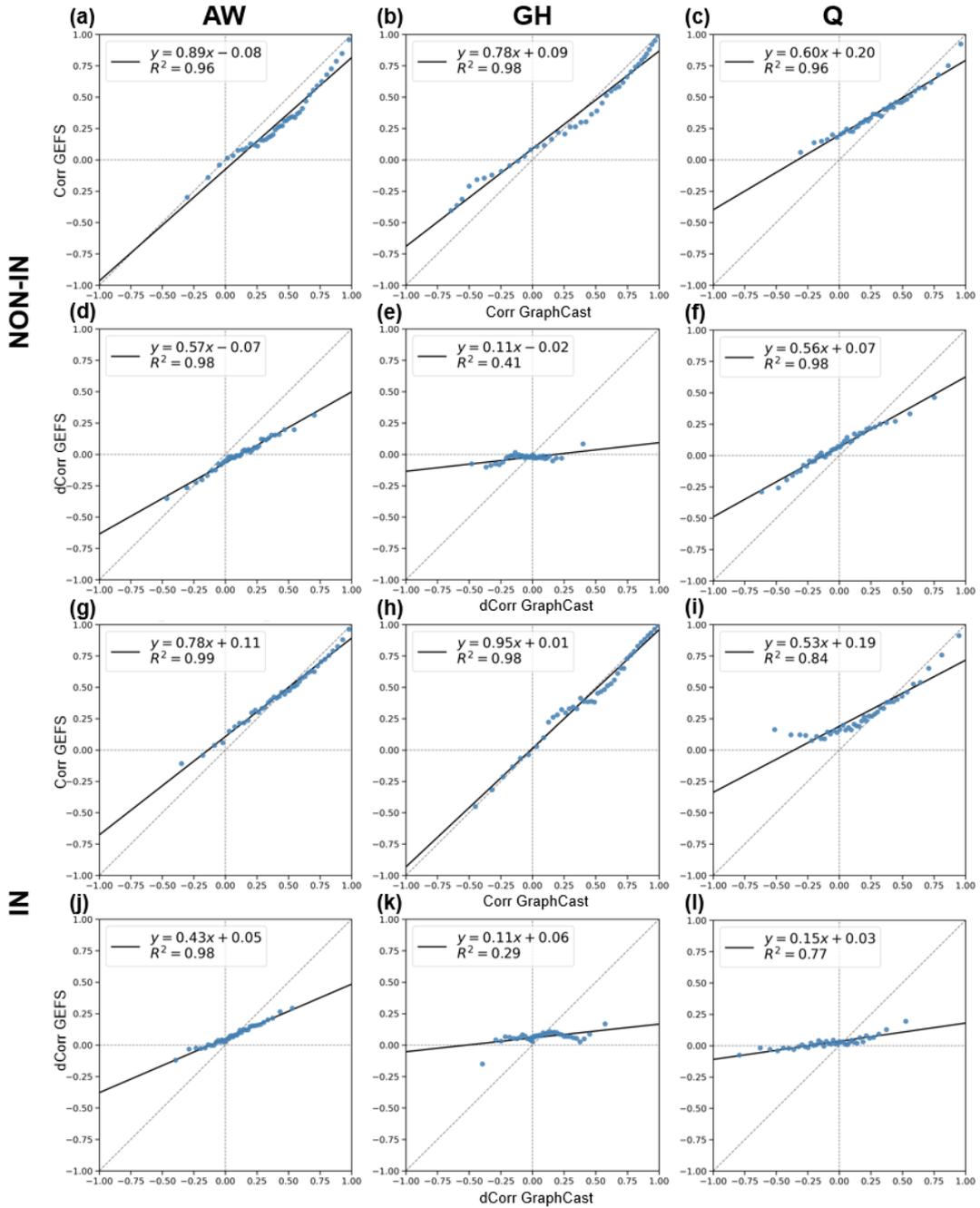

**Figure 4**. Binned scatter plots comparing GraphCast (x-axis) and GEFS (y-axis) univariate correlations and correlation tendencies within the hurricane vortex. Columns correspond to azimuthal wind (AW; left), geopotential height (GH; center), and specific humidity (Q; right). Rows are organized as: correlation for NON-IN (a–c), correlation tendency for NON-IN (d–f), correlation for IN (g–i), and correlation tendency for IN (j–l). Blue dots are binned means computed from 40 equal-width bins of the GraphCast correlation values; for each bin, the corresponding GEFS values over the same matrix entries are averaged. The solid black line is the linear regression fit, with the equation and $R^2$ shown in the upper-left inset. The dashed gray line is y = x.

Variable-dependent differences of correlations within the hurricane vortex are further examined using the relative and absolute L2 distances between the GraphCast and GEFS correlation matrices (Fig. 5), including both univariable correlations and cross-variable correlations. In Fig. 5, the diagonal elements show smaller distances than most entries in their corresponding rows and columns, suggesting the two models agree more on univariable correlations than cross-variable correlations. This is consistent with previous DA studies, which found that the cross-variable correlations in hurricanes are harder to estimate and more sensitive to sampling and model error than univariable correlations (e.g. Aksoy et al., 2012; Zhang and Weng, 2012). Among all rows, RW exhibits the largest distances, indicating that the correlations involving secondary circulation differ the most between GraphCast and GEFS. This is, again, consistent with the previous studies that unbalanced and diabatic processes generate faster, less predictable error growth (Zhang et al., 2007).

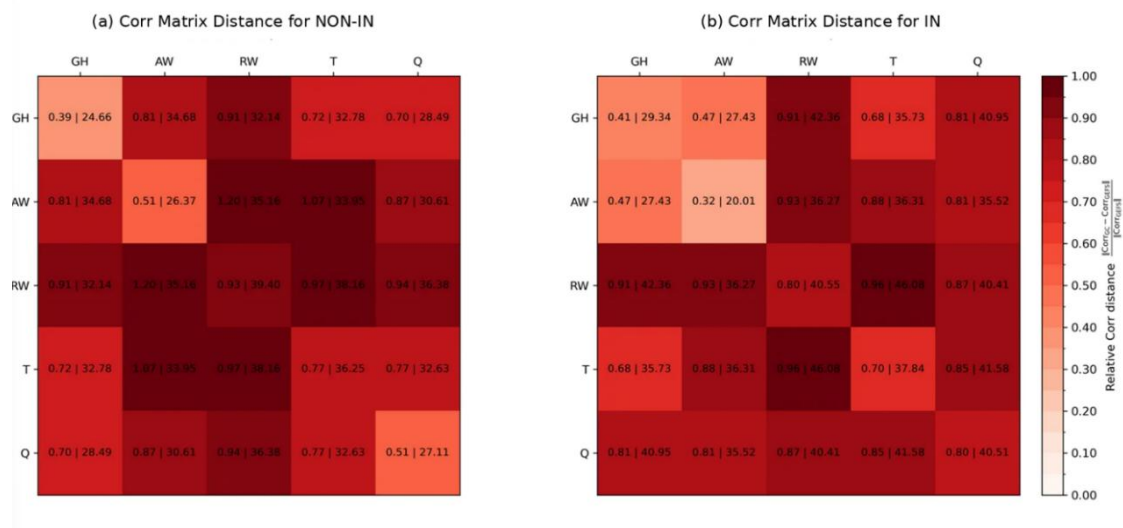

**Figure 5.** Relative (color shading and text) and absolute (text) L2 distances between GC and GEFS correlation matrices within the hurricane vortex for the NON-IN (a) and IN (b) phases. Diagonal cells represent univariate correlations; off-diagonal cells represent cross-variable correlations. Lighter shading indicates closer agreement between the two ensembles.

Motivated by the matric difference in RW, we further examine the correlation structure between the 850 hPa GH at the cyclone center and the RW. A decrease in 850 hPa GH should hypothetically be correlated with increase in upper-level outflow and lower-level inflow. For GEFS, the correlations are weak during NON-IN (Fig 6b) but become substantially stronger during IN (Fig 6d). This suggests RW is dynamically linked to the center pressure change during cyclone intensification, but this linkage is less organized if the cyclone is not intensifying. Meanwhile, in GraphCast, the correlation is stronger during NON-IN compared to GEFS (Fig 6a). The magnitude of correlation increases slightly in IN (Fig 6c), with weaker upper-level outflow and stronger lower-level inflow correlations compared to GEFS. This result suggests the dynamic link between cyclone center pressure and radial wind is less distinguished by GraphCast regardless of intensification or not, which could be attributed to the smoothing effect induced during training. Moreover, GraphCast highlights the positive correlation between cyclone center GH and the mid-level inflow at 400-500 hPa (Fig 6a, c). The mid-level inflow below the upper-level outflow ensures a sharp vertical wind shear from anticyclonic winds to cyclonic winds, balancing the warm core centered near 300 hPa (Holland & Merrill, 1984). Due to the limitation of the case study approach, we do not infer which correlation structure is closer to reality; rather, Fig. 6 shows differences in how GraphCast and GEFS couple cyclone intensity with secondary circulation, and how the coupling changes (or does not change) between NON-IN and IN.

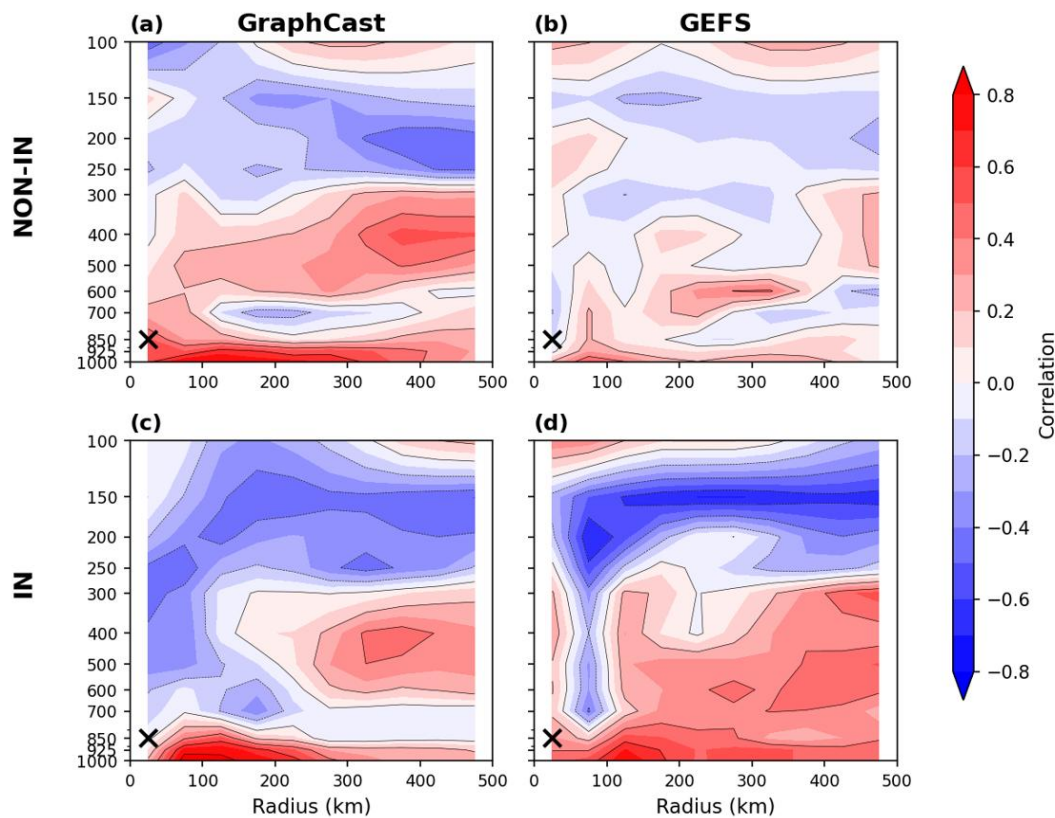

**Figure 6.** Background ensemble correlation between 850 hPa geopotential height at the cyclone center and radial wind on a radius–pressure cross section for GraphCast and GEFS during NON-IN and IN. The x-axis is radial distance from the hurricane center (0–500 km); the y-axis is pressure level (1000–100 hPa). Color shading and contours denote correlation coefficient. The black "×" marker near the surface at the cyclone center indicates the reference point (850 hPa GH at the center) against which all correlations are computed.

*b. Hurricane Environment*

While the previous section focuses on the hurricane vortex, the BEC in the large-scale environment are equally important for hurricane data assimilation and forecast. Environmental synoptic features such as the jet stream and the subtropical high are dynamically coupled; observations of one feature through data assimilation can propagate information to the others through BEC, ultimately benefiting the representation of the hurricane steering flow, wind shear, and moisture environment.

We first examine the background ensemble spread of the 500 hPa GH of the hurricane environment using Principal Component Analysis (PCA). As shown in Fig. 7, GraphCast has much smaller absolute variance for the leading EOFs. During NON-IN (Fig. 7a), the variance of the GraphCast EOF1 (950 m$^2$, the black square) is 48% of that of GEFS (around 2000 m$^2$, the red circle). During IN, while the absolute variance grows compared to NON-IN for both ensembles (1200 m$^2$ for GraphCast and 2900 m$^2$ for GEFS), the variance of the GraphCast is around 41% of that of GEFS. For the variance explained, the leading modes of GraphCast explain less cumulative variance than GEFS (e.g., top 10: 42.4% vs 45.7% during IN), showing a flatter spectrum for both phases. The above results suggest that not only does GraphCast sample a smaller spread in the hurricane environment, but it also distributes the variance across more modes with less dominance of leading patterns.

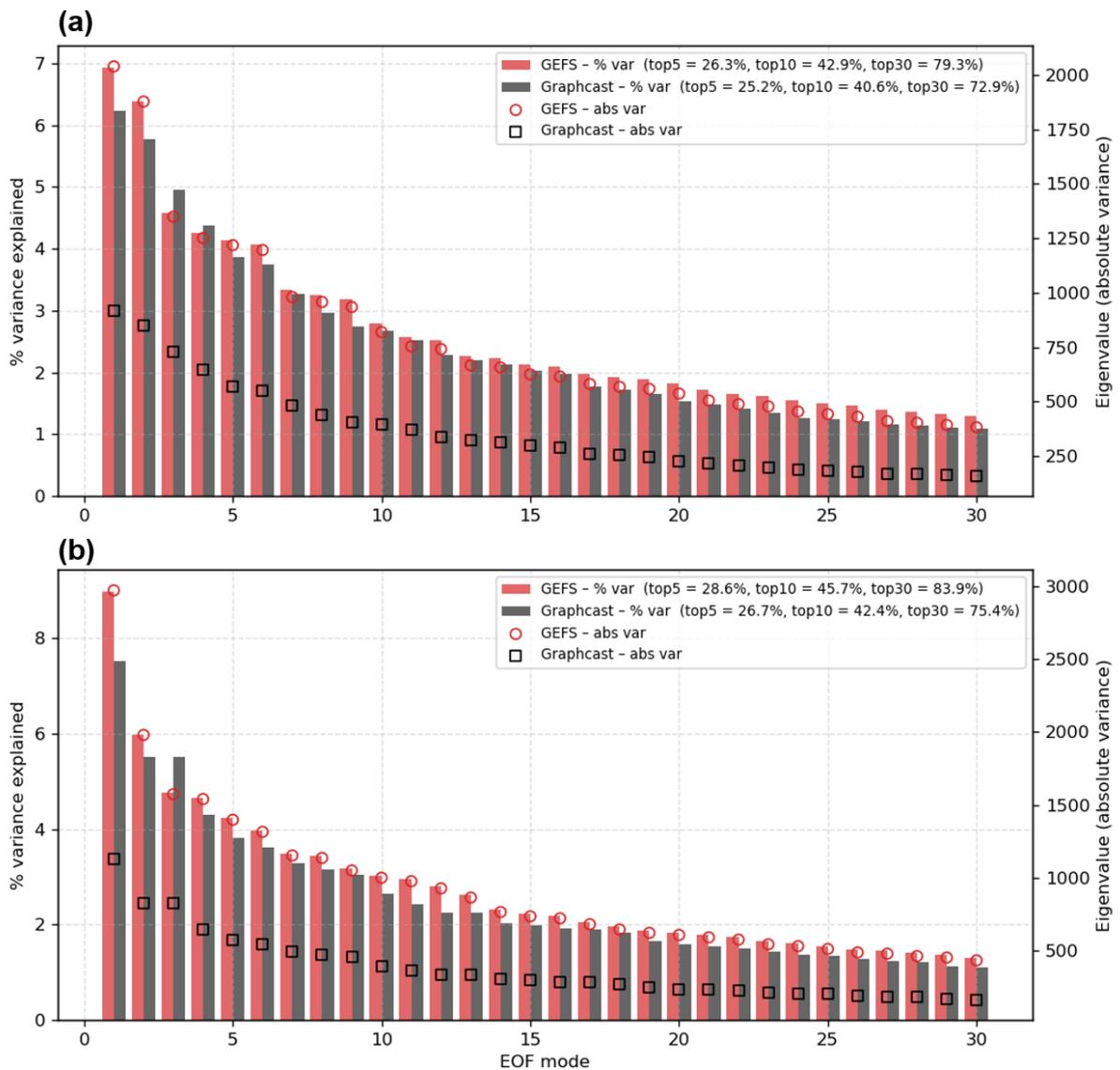

**Figure 7.** EOF spectra of the 500 hPa geopotential height background ensemble covariance over the hurricane environment domain for (a) NON-IN and (b) IN. Bars show the percentage of variance explained by each EOF mode (left y-axis), red bars for GEFS and gray bars for GraphCast. Markers show the absolute variance (eigenvalue; right y-axis, in m²): red open circles for GEFS and black open squares for GraphCast. The legend lists cumulative variance explained by the top 5, top 10, and top 30 modes for each ensemble.

To diagnose how the models distribute variance across leading modes, the spatial patterns of the leading EOFs of 500 hPa GH are examined. For both phases, the first two EOFs of GraphCast closely resemble those of GEFS, capturing the dominant, basin-scale wave pattern and jet-entrained ridge–trough patterns. Divergences emerge in less-leading modes. In NON-IN, the two ensembles begin to differ from EOF3 and onward. GEFS retains large-scale wave patterns, whereas GraphCast shows modes of short waves along the jet (Fig. 8 c, d). In IN, agreement of the EOF spatial patterns in general persists through EOF4, although for GraphCast the variance concentrates on the newly formed hurricane to the east of Lee (Fig. 8 i, k, l). These EOF spatial pattern differences are consistent with the flatter GraphCast spectrum in Fig. 7, as GraphCast tends to distribute more variance into smaller scale features such as the shortwaves, the newly formed cyclone, etc., compared to GEFS. The EOF analysis suggests that GraphCast BEC may require scale-dependent tuning of localization and inflation to maintain adequate background ensemble spread and a proper variance distribution.

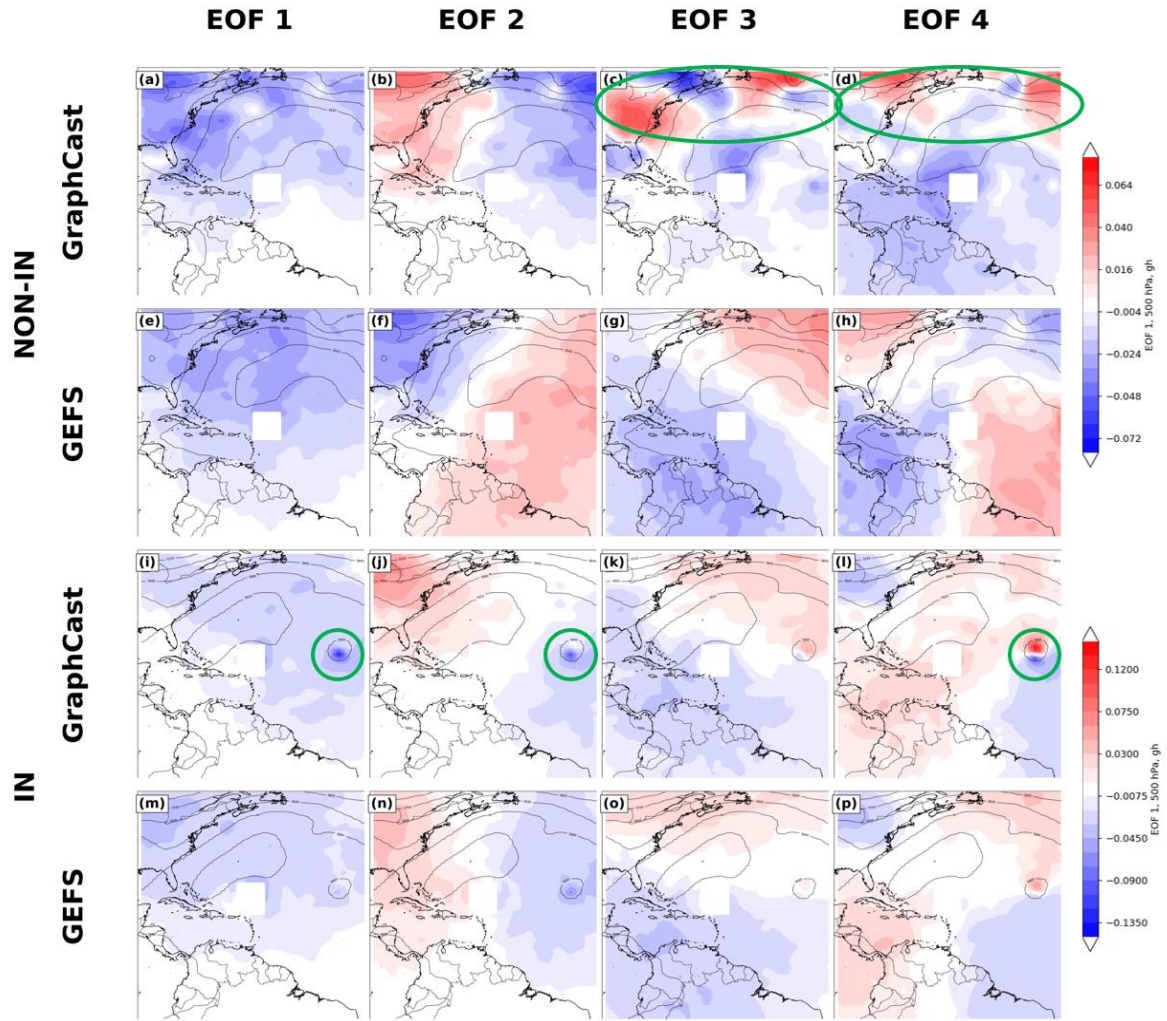

**Figure 8.** Spatial patterns of leading EOFs of 500 hPa GH BEC over the hurricane environment domain (defined in Section 2). Rows are: GraphCast NON-IN (a–d), GEFS NON-IN (e–h), GraphCast IN (i–l), and GEFS IN (m–p). Color shading shows the normalized EOF vector with separate colorbars for NON-IN and IN. of GC (a-d, i-l) and GEFS (e-h, m-p) for NON-IN (a-h) and IN (i-p). Black contours show the ensemble mean 500 hPa geopotential height (m) of either Graphcast or GEFS. Green circles highlight features discussed in the text: large circles in (c, d) highlight shortwave patterns along the jet in GraphCast NON-IN; small circles in (j, k, l) highlight the newly formed hurricane east of Lee in GraphCast IN.

Besides the 500 hPa GH discussed above, the variable dependence of correlation and correlation tendency for the hurricane environment is also examined. The binned linear regression of correlation and correlation tendency shows linear relationships for every variable between GraphCast and GEFS ($R^2 > 0.95$), demonstrating a broad consistency in how the two short-term background ensembles represent environmental correlations and their short-term evolution (T and GH are shown as example, Fig. 9). However, in the perspective of the probability density distribution, while the consistency between GraphCast and GEFS still holds for T, Q, and wind, difference is found in GH. For two grids not too far from each other, the 500 hPa geopotential heights of the two grids are likely to be positive correlated, which explains why the PDFs in Fig 9a and 9d peak at a positive value. GraphCast's GH correlation tendency PDF (Fig 9 b,e, black curve) peaks at the negative value while GEFS' (red curve) peaks at a positive value, and the resulting correlation PDF of GraphCast peaks at a smaller positive value (Fig 9 a,d), suggesting grids are less correlated with each other in GH in GraphCast compared to GEFS. A possible explanation is that large-scale mass-field adjustments in the hurricane environment are represented differently in the data-driven model, given the FV3 dynamics core for GEFS enforces mass conservation (Putman & Lin 2007) whereas GraphCast does not.

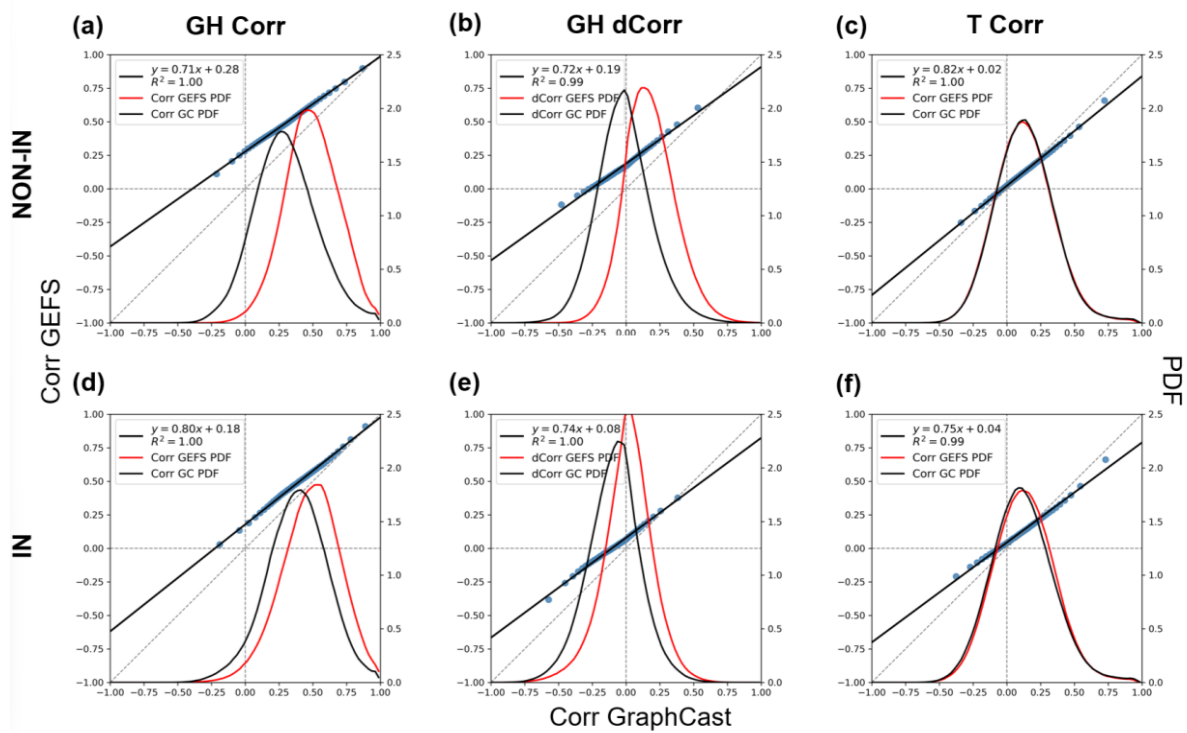

**Figure 9.** Binned scatter plots and probability density functions (PDFs) comparing GraphCast (x-axis) and GEFS (y-axis) correlation tendencies over the hurricane environment domain with a 3000 km cutoff radius. Columns correspond to GH correlation (left), GH correlation tendency (center), and temperature correlation (right). Rows are NON-IN (a–c) and IN (d–f). Blue dots are binned means. The solid black line is the linear regression fit with equation and $R^2$ shown in the inset. The dashed gray line is $y = x$. Overlaid curves show the PDFs (right y-axis): black for GraphCast, red for GEFS.

*c. Vortex-environment Interaction*

Hurricane evolution depends not only on the inner-core vortex and the surrounding synoptic flow but also on how they dynamically couple through shear, moisture transport, and outflow–environment interactions, we next examine the vortex–environment background ensemble correlation structure to assess whether GraphCast and GEFS represent these cross-scale linkages consistently. Background ensemble correlations are computed between the 500 hPa GH and two metrics, the minimum central sea-level pressure (SLP) and the longitude of the minimum SLP. Longitude is used because hurricane movement is predominantly westward in both phases.

During NON-IN (Fig. 10 a, b), GraphCast and GEFS show similar correlation patterns between minimum central SLP and 500-hPa height around the subtropical high (blue circle), the shortwave ridge over the western Atlantic (green circle), and the distant jet over the North Atlantic (purple circle). However, GEFS shows additional statistically significant positive correlations at the tropical region in the western Atlantic (red circle), where GraphCast also shows positive correlation, but the correlation is not statistically significant. During IN (Fig. 10 c,d), the two models differ more. GEFS highlights the positive correlation at the subtropical high, which is consistent with established mechanisms: strong mid-level vertical wind shear induced by the subtropical high disrupts the deep convection and weakens the hurricane (Wong and Chan, 2004). GraphCast instead highlights the statistically significant correlation between Lee's intensity and the southwest shift of the newly formed hurricane (red circle), suggesting Lee can modify the downstream large-scale circulation and modulate the steering and development of the downstream hurricane (Archambault et al., 2013, Prince and Evans, 2020).

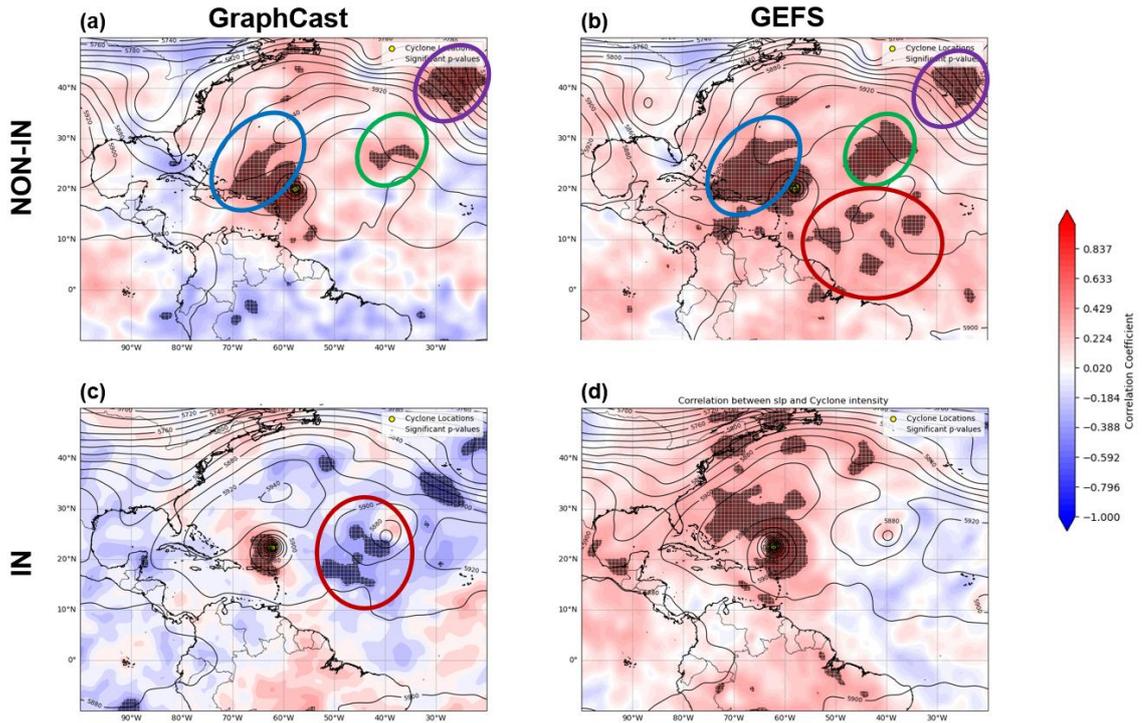

**Figure 10.** Background ensemble correlation between minimum central sea-level pressure (SLP) and 500 hPa geopotential height of the 6-hour forecast over the hurricane environment domain (including the vortex region) for GraphCast (a, c) and GEFS (b, d) during NON-IN (a, b) and IN (c, d). Color shading denotes the correlation coefficient. Hatched areas with cross markers denote grid points where the correlation is statistically significant at $p < 0.05$. Black contours show the ensemble mean 500 hPa geopotential height (m). Colored circles highlight synoptic features referenced in the text: blue circles denote the subtropical high, green circles denote the shortwave ridge over the western Atlantic, purple circles denote the jet over the North Atlantic, and red circles denote the tropical western Atlantic region during NON-IN and the newly formed hurricane during IN.

For the correlation between the longitude of the minimum SLP and 500 hPa GH (Fig 12), GraphCast and GEFS agree with each other with relatively minor differences. During NON-IN, both models show the statistically significant negative signal over the subtropical high, which reflects the impact of the westward steering flow associated with the subtropical high, while GraphCast finds several positive signals in the tropics. During IN, both models agree on the negative correlation at the subtropical high, while GEFS shows a larger area of statistically significant negative signal at the ITCZ. GraphCast and GEFS agree more on the environment-track coupling than on the environment-intensity coupling, which is physically expected because hurricane motion is dominated by deep-layer steering flow of the environment, while intensity is strongly controlled by inner-core deep convections in the 6-hour short lead time (Emanuel and Zhang, 2017).

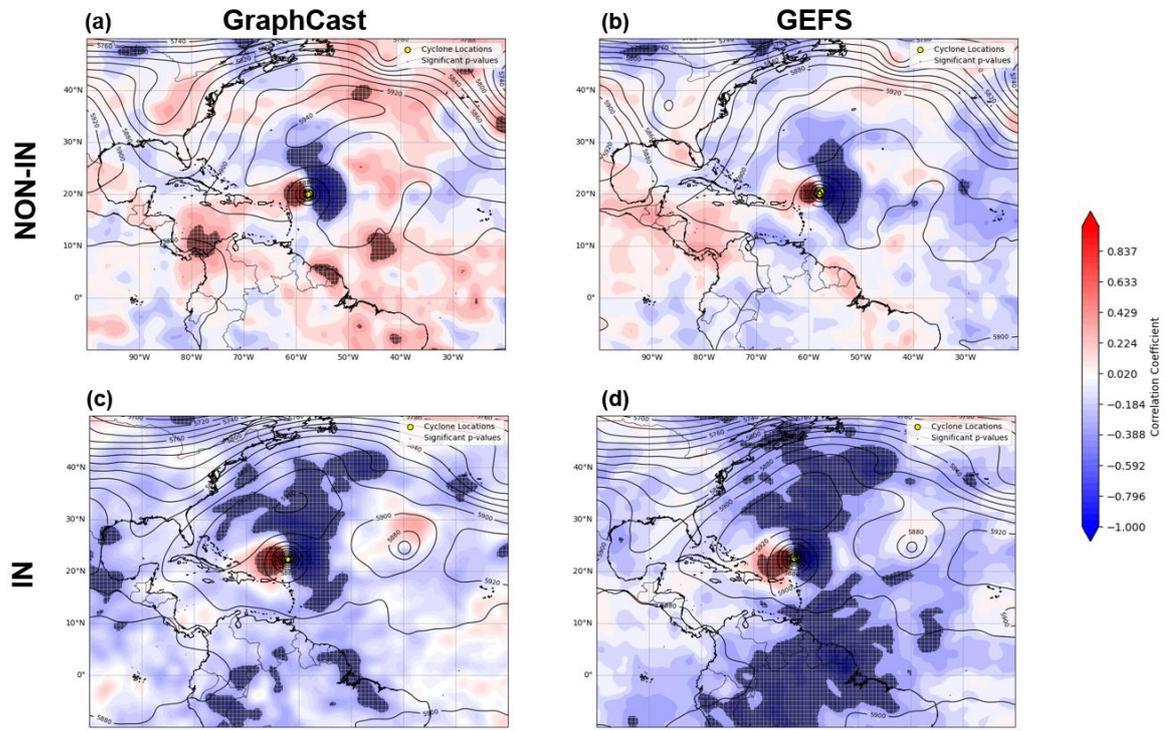

**Figure 11.** Same as Fig. 10, but for the background ensemble correlation between the longitude of the minimum SLP and 500 hPa geopotential height.

## 4. Discussions and Conclusions

This study evaluates whether GraphCast can produce short-term BEC with realistic amplitude and structure for hurricane data assimilation. The 6-hour BEC estimated from GraphCast are compared to those from GEFS for Hurricane Lee (2023) during both a non-intensification and an intensification phase. The key findings are summarized below, organized by the hurricane vortex, the hurricane environment, and hurricane vortex–environment interactions, followed by discussion on geopotential height, limitations, and the implications of this study for future ensemble-based DA.

Within the hurricane vortex, GraphCast shows smaller background ensemble variance (Fig. 2), similar magnitudes of correlation, and larger magnitudes of correlation tendency (Fig. 3) compared to GEFS. The two models agree well in background ensemble correlations for balanced processes—azimuthal wind, which describes the primary circulation tied to gradient-wind and hydrostatic balance, showing consistent correlations and correlation tendencies between GraphCast and GEFS (Fig. 4). The two models diverge for unbalanced processes—radial wind, which is linked to the secondary circulation driven by boundary-layer friction and deep convection, showing the largest correlation differences (Fig. 5, 6); moisture correlations and correlation tendencies also diverge during IN (Fig. 4). Cross-variable correlation differences between GraphCast and GEFS are larger than univariate correlation differences (Fig. 5).

Within the hurricane environment, GraphCast also shows smaller background ensemble variance than GEFS (Fig. 7), consistent with the under-dispersion issue found within the hurricane vortex. GraphCast also shows a flatter EOF spectrum (Fig. 7) compared to GEFS, indicating that the background ensemble perturbations are less concentrated in the leading modes and more dispersed among less dominating modes. Further diagnosis reveals that in the leading EOFs, GraphCast distributes more variance to sub-synoptic scale features such as shortwaves embedded in the jet stream and the newly formed hurricane (Fig. 8). The background ensemble correlations and correlation tendencies of most variables show linear relationships between GC and GEFS after binning (Fig. 9), indicating that GraphCast produces broadly matching correlation structure of the hurricane environment. The one notable exception is GH: the probability density function of GH correlations in GraphCast peaks at smaller values than in GEFS (Fig. 9), indicating that GH within the hurricane environment is, on average, less spatially correlated in GraphCast than in GEFS.

For the hurricane vortex–environment interactions, the two models produce background ensemble correlation patterns for cyclone track with relatively minor mismatch than the intensity (Fig. 10 and Fig. 11). For cyclone intensity, the two models agree with statistical significance on the synoptic features during NON-IN. However, during IN, GraphCast highlights the correlation with the newly formed hurricane, while GEFS emphasizes the influence of the subtropical high (Fig. 10), suggesting that the two models identify different large-scale mechanisms affecting the intensification of Hurricane Lee.

Compared to other variables, GH requires greater caution if we want to use GraphCast as an emulator in DA. Within the hurricane vortex, GH shows the largest initial correlations and the smallest correlation tendencies compared to other variables (Fig 3). While the strong correlations tied to the gradient wind and hydrostatic balance match well between two models, the weak correlation tendencies, involving complex mechanisms such as latent heating release and mass transport, show little agreement (Fig.4). In the hurricane environment, geopotential height is less correlated spatially in GraphCast than in GEFS (Fig. 9), which is likely due to the lack of the explicit mass conservation constraints during the GraphCast training. Degradations in GH have also been found in an end-to-end data-driven weather forecasting system with cycling DA (Ni et al., 2025). Efforts have been made to improve the mass balance in data driven models by changing the weight of vertical layers in the loss function from pressure-dependent to density-dependent (Wang et al., 2024).

As an initial step of exploring the GraphCast fidelity in surrogating the background ensemble for potential hurricane DA, a single case is selected to allow in-depth diagnostics. Cautions are warranted to extend the results to other cases. To facilitate a controlled comparison of the BECs, isolating the choice of model, emulator versus physics-based, as the primary factor, this study does not perform a cycling DA experiment. Instead, this experiment design enforces that both systems share identical initial perturbations from the GEFS analysis. A recent study using a two-layer quasi-geostrophic model shows that a hybrid ensemble DA system consisting of both a physics-based ensemble and a data-driven ensemble outperforms DA systems with either ensemble alone (Kubalek et al 2026). To further explore the potential of data-driven models as background ensemble surrogates in ensemble DA, we plan to construct a hybrid ensemble DA system using JEDI, combining GraphCast and GEFS in background ensemble construction, and to evaluate the hybrid system with cycling DA experiments focusing on hurricane analysis and forecast.

*Acknowledgments*

This work is supported by NOAA Grant NA22NWS4680011. Computing for this project was performed at the OU Supercomputing Center for Education and Research (OSCER) at the University of Oklahoma (OU).

*Data Availability Statement*

The GEFS analysis and forecast data used in this study are available from the AWS Open Data Registry for GEFS (https://noaa-gefs-pds.s3.amazonaws.com) provided by the NOAA Open Data Dissemination Program. GraphCast source code and pretrained weights are publicly available through DeepMind's GitHub repository (https://github.com/google-deepmind/graphcast).